# Robust synchronization with uniform ultimate bound between two different chaotic systems with uncertainties

Jianping Cai[*] , Zhengzhong Yuan
Department of Mathematics and Information Science,
Zhangzhou Normal University, Zhangzhou 363000, China
[*] mathcai@hotmail.com

**Abstract** Adaptive controllers are designed to synchronize two different chaotic systems with uncertainties, including unknown parameters, internal and external perturbations. Lyapunov stability theory is applied to prove that under some conditions the drive-response systems can achieve synchronization with uniform ultimate bound even though the bounds of uncertainties are not known exactly in advance. The designed controllers contain only feedback terms and partial nonlinear terms of the systems, and they are easy to implement in practice. The Lorenz system and Chen system are chosen as the illustrative example to verify the validity of the proposed method. Simulation results also show that the present control has good robustness against different kinds of disturbances.

## 1 Introduction

Beginning with the pioneering work of Pecora and Carroll,[1] synchronization of chaotic systems has become an important topic in nonlinear science not only for its importance in theory but also for its potential applications in various areas, for example, secure communication,[2,3] chemical and biomedical science,[4,5] life science,[6] electromechanical engineering[7] and so on.[8,9] Many different control strategies have been developed to deal with synchronization issue such as replacing variable control,[1] linear or nonlinear feedback control,[10-12] sliding mode control,[13] adaptive control,[14] impulsive control.[15] Synchronization of two identical chaotic systems has been widely studied, but in real-world system such as laser arrays,[16] biological system,[5] it is hardly the case that every component can be assumed to be identical. As a consequence, synchronization between two different chaotic systems has attracted increasing attention.[17,18] Transmitter and receiver with different chaotic structures imply larger key space and higher security in a cryptosystem when synchronization is applied to secure communication. Moreover, uncertainties, including unknown parameter, internal and external perturbation, occur commonly in practical situation. Such uncertainties may destroy the synchronization,[19] or achieve higher security of cryptosystems by intentionally injecting them into chaotic systems.[20] Hence, synchronization between two different chaotic systems in the presence of uncertainties is becoming an important issue. Synchronization between two different chaotic systems with unknown parameters was investigated.[21-26] While the chaotic system is perturbed by random or regular noise, synchronization issue is studied.[27-31] Adaptive control is one of the most efficient methods to deal with these problems with uncertainties.

In this paper adaptive control technique is used to design suitable controllers to synchronize different chaotic systems with uncertainties, including unknown parameters, internal and external perturbations (random or regular). Because of the existence of uncertainties, the drive-response systems are usually difficult to achieve complete synchronization. So a concept of synchronization with uniform ultimate bound is introduced in this paper. Under some conditions Lyapunov stability theory ensures that the drive-response chaotic systems can achieve synchronization with uniform ultimate bound even though the bounds of uncertainties are not known exactly in advance. The designed controllers contain only feedback terms and partial nonlinear terms of the systems, and they are easy to implement in practice. The Lorenz system and Chen system are chosen as the illustrative example to verify the validity of the proposed method. Simulation results also show that the present control is robust against different kinds of disturbances.

## 2 Problem description

Consider the drive system of the form

$$\dot{x} = f(x) + F(x)\alpha + d(x,t), \tag{1}$$

where $x \in R^n$ is the state vector, $\alpha \in R^m$ is the unknown parameter vector, $f(x) \in R^{n\times 1}$ and $F(x) \in R^{n\times m}$ are known function matrices, $d(x,t) \in R^{n\times 1}$ is the uncertainties including parameter perturbation and external disturbance.

Let $\Omega_x \subset R^n$ be a bounded region containing the whole attractor of drive system (1) such that no trajectory of system (1) ever leaves it. This assumption is simply based on the bounded property of chaotic attractor. Also, let $M \subset R^m$ be the set of parameter under which system (1) is in a chaotic state.

The response system with a controller is constructed as follows

$$\dot{y} = g(y) + G(y)\beta + u(t), \tag{2}$$

where $y \in R^n$ , $\beta \in R^s$ is the unknown parameter vector, $g(y) \in R^{n\times 1}$ and $G(y) \in R^{n\times s}$ are known function matrices, $u(t) \in R^n$ is the control input vector. Let $\Omega_y \subset R^n$ be a bounded region containing the whole attractor of response system (2) with $u(t) = 0$.

Due to the uncertainties in the drive system (1), the response system (2) is usually difficult to achieve complete synchronization with the drive system (1). Therefore a concept of synchronization with uniform ultimate bound is introduced in this paper.

**Definition 1** The drive-response systems (1) and (2) achieve synchronization with uniform ultimate bound if for any initial state $x(0)\in\Omega_x$ and $y(0)\in\Omega_y$, there exist constants $h>0$ and $T_0>0$ such that the trajectory $x(t,x(0))$ of system (1) and trajectory $y(t,y(0))$ of system (2) satisfy

$$\|x(t,x(0))-y(t,y(0))\|<h, \text{ for any } t>T_0, \tag{3}$$

where ||.|| denotes the Euclidean norm.

According to this definition, our control objective is to design a suitable adaptive controller $u(t)$ such that the distance between trajectories $x(t)$ and $y(t)$ of drive and response systems will eventually become less than an error bound $h$.

## 3 Design of adaptive synchronization controller

Firstly, some assumptions are given as follows.

**Assumption 1** The disturbance vector $d(x,t)$ are norm bounded by an unknown positive constant $L_d$, namely, $\|d(x,t)\|\le L_d$.

**Assumption 2** The function vectors $f(\cdot)$ and $g(\cdot)$ are continuous on a bounded closed region $\Omega$ containing both $\Omega_x$ and $\Omega_y$. So there exists a positive constant $L_f$ such that

$$\|f(x)-g(x)\|\le L_f,\quad x\in\Omega.$$

**Assumption 3** The function vector $g(\cdot)$ satisfies the Lipschitz condition, that is, there exists a positive constant $L_g$ such that

$$\|g(x)-g(y)\|\le L_g\|x-y\|, \text{ for any } x,y\in R^n.$$

**Theorem 1** Under Assumptions 1-3, the drive-response systems (1) and (2) can achieve synchronization with uniform ultimate bound if the controller $u(t)$ is designed as

$$u(t) = F(x)\hat{\alpha} - G(y)\hat{\beta} + \hat{k}e\,, \tag{4}$$

where $e = x - y$ is the error variable, and the adaptive variables $\hat{\alpha}$, $\hat{\beta}$ and $\hat{k}$ satisfy the following adaptation laws

$$\dot{\hat{\alpha}} = F(x)^T e\,,\ \ \dot{\hat{\beta}} = -G(y)^T e\,,\ \ \dot{\hat{k}} = \| e \|^2\,. \tag{5}$$

**Proof** The error dynamical system is

$$\dot{e} = \dot{x} - \dot{y} = f(x) - g(y) + F(x)(\alpha - \hat{\alpha}) - G(y)(\beta - \hat{\beta}) - \hat{k}e\,. \tag{6}$$

Construct a Lyapunov function

$$V = \frac{1}{2}[e^T e + (\alpha - \hat{\alpha})^T(\alpha - \hat{\alpha}) + (\beta - \hat{\beta})^T(\beta - \hat{\beta}) + (k - \hat{k})^2]\,,$$

where $k$ is a constant to be determined. Using Assumptions 1-3 and Eqs.(4)-(6), the time derivative of $V$ satisfies,

$$\begin{aligned}
\dot{V} &= e^T\dot{e} - (\alpha - \hat{\alpha})^T\dot{\hat{\alpha}} - (\beta - \hat{\beta})^T\dot{\hat{\beta}} - (k - \hat{k})\dot{\hat{k}} \\
&= e^T(f(x) - g(y) + F(x)(\alpha - \hat{\alpha}) - G(y)(\beta - \hat{\beta}) - \hat{k}e + d(x,t)) \\
&\quad - (\alpha - \hat{\alpha})^T F(x)^T e - (\beta - \hat{\beta})^T(-G(y)^T e) - (k - \hat{k})\| e \|^2 \\
&= e^T(f(x) - g(x)) + e^T(g(x) - g(y)) - k\| e \|^2 + e^T d(x,t) \\
&\le L_f \| e \| + L_g \| e \|^2 - k\| e \|^2 + L_d \| e \| \\
&\le \frac{\| e \|^2}{2\varepsilon_1} + \frac{\varepsilon_1 L_f^2}{2} + L_g \| e \|^2 - k\| e \|^2 + \frac{\| e \|^2}{2\varepsilon_2} + \frac{\varepsilon_2 L_d^2}{2} \\
&= (\frac{1}{2\varepsilon_1} + \frac{1}{2\varepsilon_2} + L_g - k)\| e \|^2 + \frac{1}{2}(\varepsilon_1 L_f^2 + \varepsilon_2 L_d^2)
\end{aligned}$$

where $\varepsilon_1$, $\varepsilon_2$ are small positive constants. Let

$$\varepsilon = \frac{1}{2}(\varepsilon_1 L_f^2 + \varepsilon_2 L_d^2)\,. \tag{7}$$

If the constant $k$ satisfies

$$k \ge \frac{1}{2\varepsilon_1} + \frac{1}{2\varepsilon_2} + L_g + 1\,, \tag{8}$$

then we have

$$\dot{V} \le -\| e \|^2 + \varepsilon\,. \tag{9}$$

Based on (9), synchronization with uniform ultimate bound thus follows using the results and terminology.[32,33] Therefore, the state error will be contained within the vicinity of the equilibrium $e=0$.

**Remark 1** From Eqs.(7) and (9), it is clear that if $\varepsilon_1$ and $\varepsilon_2$ are chosen sufficiently small, $\varepsilon$ will be sufficiently small which implies the synchronization error will also be sufficiently small.

**Remark 2** The controller (4) contains only the feedback term and partial nonlinear terms of the systems, while the controllers[22-28] include all the information appeared in the error dynamical system or the master-slave systems.

**Remark 3** Our method needn't know the exact bounds of state variables $y_i$ of the response system, which are hard to determine in practice. But the bounds must be known in implementation.[29]

**Remark 4** From Eqs.(7)-(9), not only the bounds $L_d$ and $L_f$, but also the Lipschitz constant $L_g$ can be injected into the constants $k$ and $\varepsilon$, which can be adaptively adjusted by the adaptation laws (5). So the constants $L_d$, $L_f$ and $L_g$ are necessary in the process of theoretical proof but unnecessary to be known exactly in practice.

**Remark 5** As we will see in what follows, the present control is robust against different kinds of disturbances, including internal or external, random or regular disturbances.

## 4 Illustrative example

Synchronization between Lorenz system and Chen system is presented to simulate the proposed method.

The Lorenz system with unknown parameters and perturbations is chosen as the drive system

$$\begin{pmatrix} \dot{x}_1 \\ \dot{x}_2 \\ \dot{x}_3 \end{pmatrix} = \begin{pmatrix} 0 \\ -x_2 - x_1 x_3 \\ x_1 x_2 \end{pmatrix} + \begin{pmatrix} x_2 - x_1 & 0 & 0 \\ 0 & x_1 & 0 \\ 0 & 0 & -x_3 \end{pmatrix} \begin{pmatrix} \alpha_1 \\ \alpha_2 \\ \alpha_3 \end{pmatrix} + \begin{pmatrix} d_1(x,t) \\ d_2(x,t) \\ d_3(x,t) \end{pmatrix} \tag{10}$$

Compared with Eq.(1), the relative notations are

$x=(x_1,x_2,x_3)^T$ , $f(x)=(0,-x_2-x_1x_3,x_1x_2)^T$ , $F(x)=diag(x_2-x_1,x_1,-x_3)$ ,

$\alpha=(\alpha_1,\alpha_2,\alpha_3)^T$, $d(x,t)=(d_1(x,t),d_2(x,t),d_3(x,t))^T$.

The Chen system with unknown parameters and controllers is selected as the response system

$$\begin{pmatrix}\dot{y}_1\\ \dot{y}_2\\ \dot{y}_3\end{pmatrix}=\begin{pmatrix}0\\ -y_1y_3\\ y_1y_2\end{pmatrix}+\begin{pmatrix}y_2-y_1 & 0 & 0\\ -y_1 & y_1+y_2 & 0\\ 0 & 0 & -y_3\end{pmatrix}\begin{pmatrix}\beta_1\\ \beta_2\\ \beta_3\end{pmatrix}+\begin{pmatrix}u_1(t)\\ u_2(t)\\ u_3(t)\end{pmatrix} \tag{11}$$

Compared with Eq.(2), the relative notations are

$$y=(y_1,y_2,y_3)^T,\ g(y)=(0,-y_1y_3,y_1y_2)^T,\ G(y)=\begin{pmatrix}y_2-y_1 & 0 & 0\\ -y_1 & y_1+y_2 & 0\\ 0 & 0 & -y_3\end{pmatrix},$$

$\beta=(\beta_1,\beta_2,\beta_3)^T$. $u(t)=(u_1(t),u_2(t),u_3(t))^T$ is determined by Eqs.(4) and (5).

It is easy to verify that the Lorenz system and Chen system satisfy Assumptions 1-3.

**Case 1** If no disturbance is applied to the drive system, that is, $d(x,t)=0$, then the response system (11) can achieve complete synchronization with the drive system (10) as shown in Fig.1, where the unknown parameters are assumed to be "known" as $\alpha=(10,28,8/3)^T$ and $\beta=(35,28,3)^T$, and the initial values are $x(0)=(3,-2,5)^T$, $y(0)=(-1,2,1)^T$, $\hat{\alpha}(0)=(3,-2,-1)^T$, $\hat{\beta}(0)=(-5,4,2)^T$ and $\hat{k}(0)=1$.

**Case 2** If parameter perturbation is applied to the drive system (10), for example,

$$\dot{x}_1=10x_2-10(1-c\sin 2t)x_1,$$
$$\dot{x}_2=28x_1-(1-c\cos t)x_2-x_1x_3,$$
$$\dot{x}_3=x_1x_2-\frac{8}{3}(1-c\sin 3t)x_3,$$

that is,

$$d(x,t)=(c\sin(2t)\cdot(10x_1),c\cos t\cdot x_2,c\sin(3t)\cdot(\frac{8}{3}x_3))^T, \tag{12}$$

the response system (11) can achieve synchronization with uniform ultimate bound with the drive system (10). Fig.2 displays the result with the perturbed strength $c=5\%$ compared with the magnitude of state variables $x_i(i=1,2,3)$. Such perturbation does not destroy the chaotic characteristic of the Lorenz system, whose attractor is shown in Fig.3. As shown in Fig.4, the proposed control is robust against parameter perturbation even though the perturbed strength $c$ increases to 10%. The ultimate synchronization error bound is about 0.27 for $c = 10\%$ while the value is 0.17 for $c = 5\%$. The adaptive processes of parameters $\hat{\alpha}(t)$, $\hat{\beta}(t)$ and $\hat{k}(t)$ are sketched in Figs.5 and 6. The initial values are the same as those of Case 1.

**Case 3** If external perturbation is applied to the drive system (10), for example,

$$d(x,t) = (cm_1 \sin(2t), cm_2 \cos t, cm_3 \sin(3t))^T, \tag{13}$$

where $m_i (i = 1,2,3)$ are respectively the bounds $x_i (i = 1,2,3)$ of attractor of the drive system (10), the response system (11) can achieve synchronization with uniform ultimate bound with the drive system (10). The simulation result is shown in Fig.7 with the perturbed strength $c = 5\%$. We can assume that $m_1 = 20$, $m_2 = 25$ and $m_3 = 50$ from Fig.3. Additional simulations verify that the ultimate synchronization error bound is about 0.09 for $c = 10\%$ while the value is 0.05 for $c = 5\%$. The above simulations are performed in computer algebraic system *Mathematica*, where instruction *NDSolve* is used.

**Case 4** If external random noise is applied to the drive system (10), for example,

$$d(x,t) = (cm_1 N(0,1), cm_2 U[0,1], cm_3 \sin t))^T, \tag{14}$$

where $N(0,1)$ is the standard normal distribution and $U[0,1]$ is the uniform distribution on [0,1], the response system (11) also can achieve synchronization with the drive system (10). The result is sketched in Fig.8, where the perturbed strength $c = 5\%$. Additional simulations show that the average of the ultimate synchronization error bounds is about 0.15 for random noise strength $c = 10\%$ while the value is 0.07 for $c = 5\%$. The simulations of this case are performed in computer algebraic system *Matlab*, where instruction *ode45* with step size 0.001 is chosen.

## 5 Conclusions

In this paper we design adaptive controllers to synchronize two different chaotic systems with uncertainties, containing unknown parameters, internal and external perturbations (regular or random). Using Lyapunov stability theory, we demonstrate that under some conditions the drive-response systems can achieve synchronization with uniform ultimate bound. There is no need to know exactly the bounds of uncertainties and the Lipchitz constant in advance. The designed controllers contain only feedback terms and partial nonlinear terms of the systems, and they are easy to implement in practice. The Lorenz system and Chen system are chosen as the illustrative example to verify the validity of the proposed method. Simulations show that the present control has good robustness against different kinds of disturbances.

**Acknowledgements** Research is supported by the National Natural Science Foundation of China under grant No 60674049.

## References

[1] L.M. Pecora and T.L. Carroll, *Phys. Rev. Lett.* **64**(1990)821.

[2] T.L. Carroll and L.M. Pecora, *IEEE Trans. Circ. Syst. II* **40** (1993) 646.

[3] L.J. Kocarev, K.S. Halle, K. Eckert, L.O. Chua and U. Parlitz, *Int. J. Bifur. Chaos* **2**(1992)709.

[4] Y. Kuramoto, *Chemical oscillations, waves and turbulence*, Springer, Berlin, 1980.

[5] A.T. Winfree, *The Geometry of biological time*, Springer, New York, 1980.

[6] L.Glass, *Nature* **410** (2001) 277.

[7] R.Yamapi and P. Woafo, *J. Sound Vib*. **285** (2005) 1151.

[8] S. Boccaletti, J. Kurths, G. Osipov, D.L. Valladares and C.S. Zhou, *Phys. Rep.* 366(2002)1.

[9] A. Pikovski, M. Rosenblum and J. Kurths, *Synchronization: A universal concept in nonlinear sciences*, Cambridge University Press, Cambridge, 2001.

[10] M.T. Yassen, *Chaos Solitons Fractals* **26**(2005)913.

[11] M. Chen and Z. Han, *Chaos Solitons Fractals* **17**(2003)709.

[12] J.P. Cai, X.F. Wu and S.H. Chen, *Phys. Scr.* **75** (2007) 379.

[13] H. Zhang, X.K. Ma and W.Z. Liu, *Chaos Solitons Fractals* **21**(2004)1249.

[14] J.H. Lv and S.H. Chen, *Chaos Solitons Fractals* **14**(2002)643.

[15] Y. Tao and L.O. Chua, *IEEE Trans. Circ. Syst. I,* **44**(1997) 976.

[16] K. Otsuka, *Phys. Rev. Lett.* **65**(1990)329.

[17] L. Huang, R. Feng and M. Wang, *Phys. Lett. A* **320**(2004)271.

[18] J.H. Park, *Chaos Solitons Fractals* **27**(2006)549.

[19] G.R. Chen and X. Dong, *From Chaos to Order: Methodologies, Perspectives, and Applications*, World Scientific, Singapore, 1998.

[20] R. Raoufi and Z. Zinober, *Int. J. Systems Science* **38**(2007) 931.

[21] R.H. Li, W. Xu and S. Li, *Phys. Lett. A* **367**(2007)199.

[22] H.G. Zhang, W. Huang, Z.L. Wang and T.Y. Chai, *Phys. Lett. A* **350**(2006)363.

[23] X.Y. Chen and J.F. Lu, *Phys. Lett. A* **364**(2007)123.

[24] J. Huang, *Phys. Lett. A* **372**(2008)4799.

[25] W. Xu, L.X. Yang and Z.K. Sun, *Nonlinear Dyn.* **52**(2008)19.

[26] R. X. Zhang and S.P. Yang, *Chin. Phys.* B **17(** 2008) 4073

[27] X.Y. Wang and M.J. Wu, *Int. J. Modern Phys. B* **22**(2008)4069.

[28] X.C. Li, W. Xu and Y.Z. Xiao, *J. Sound Vib.* **314**(2008)526.

[29] S. Bowong and J.J. Tewa, *Nonlinear Dyn.* doi: 10.1007/s11071-008-9379-6.

[30] Y.H. Sun, J.D. Cao and G. Feng, *Phys. Lett. A* **372**(2008)5442.

[31] F.L. Jia, W. Xu and L. Du, *Chinese Phys.* **16**(2007)3249.

[32] M.J. Corless and G. Leitmann, *IEEE Trans. Automatic Control* **26**(1981)1139.

[33] K.M. Koo and J.H. Kim, *IEEE Trans. Automatic Control* **39**(1994)1230.

**Figure captions**

Fig.1 Achievement of complete synchronization without perturbation

Fig.2 Achievement of synchronization with uniform ultimate bound under perturbation (12)

Fig.3 Attractor of the Lorenz system under perturbation (12): (a) $x_1 - x_2$ plane; (b) $x_1 - x_3$ plane

Fig.4 Synchronization error of drive-response systems (10) and (11) under perturbation (12) with strength $c = 10\%$

Fig.5 Adaptive processes of parameters $\hat{\alpha}_1$, $\hat{\alpha}_2$, $\hat{\alpha}_3$ and $\hat{k}$ under perturbation (12)

Fig.6 Adaptive processes of parameters $\hat{\beta}_1$, $\hat{\beta}_2$ and $\hat{\beta}_3$ under perturbation (12)

Fig.7 Achievement of synchronization with uniform ultimate bound under perturbation (13)

Fig.8 Achievement of synchronization with uniform ultimate bound under perturbation (14)

## Figures

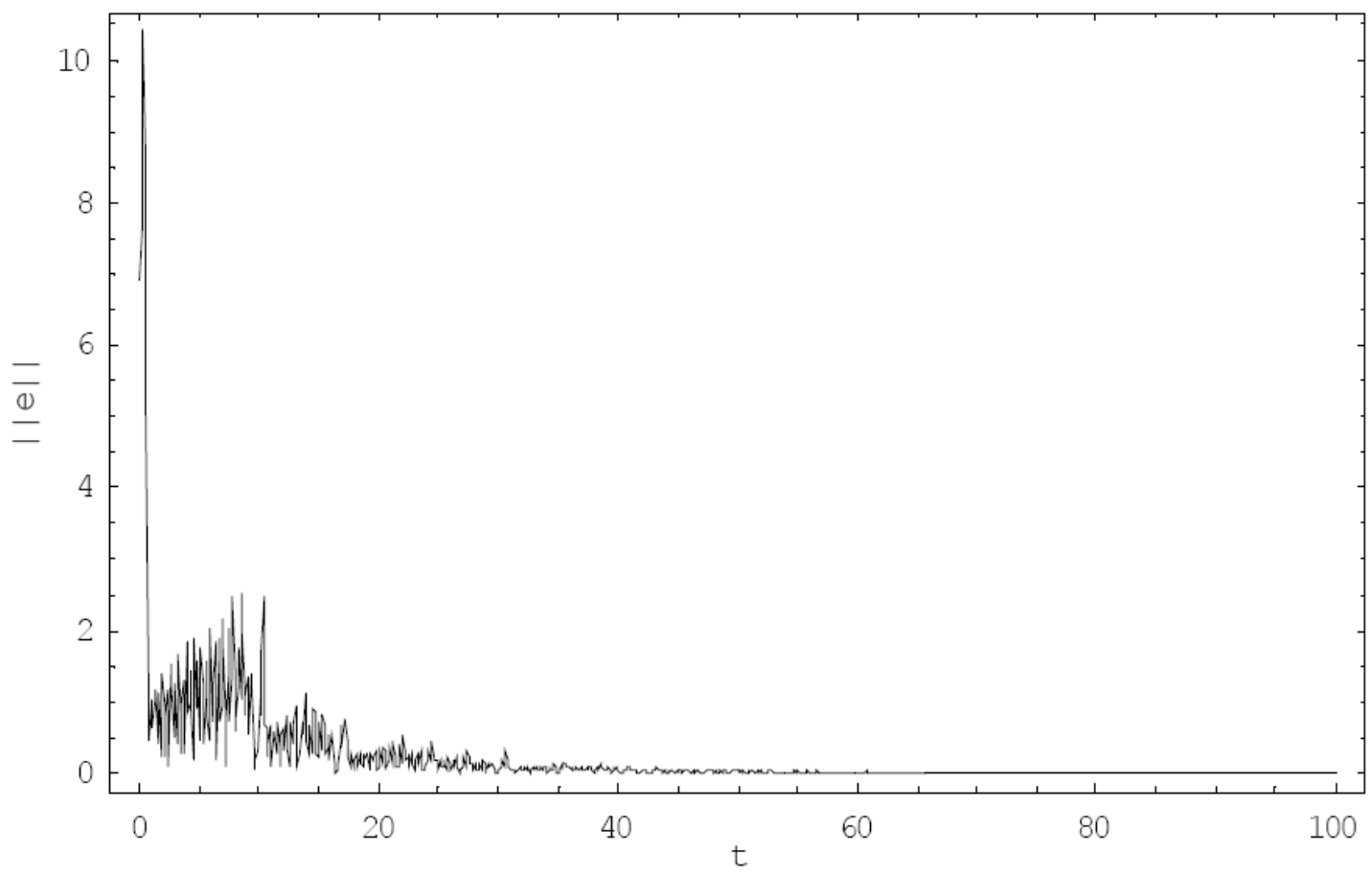

Fig.1 Achievement of complete synchronization without perturbation

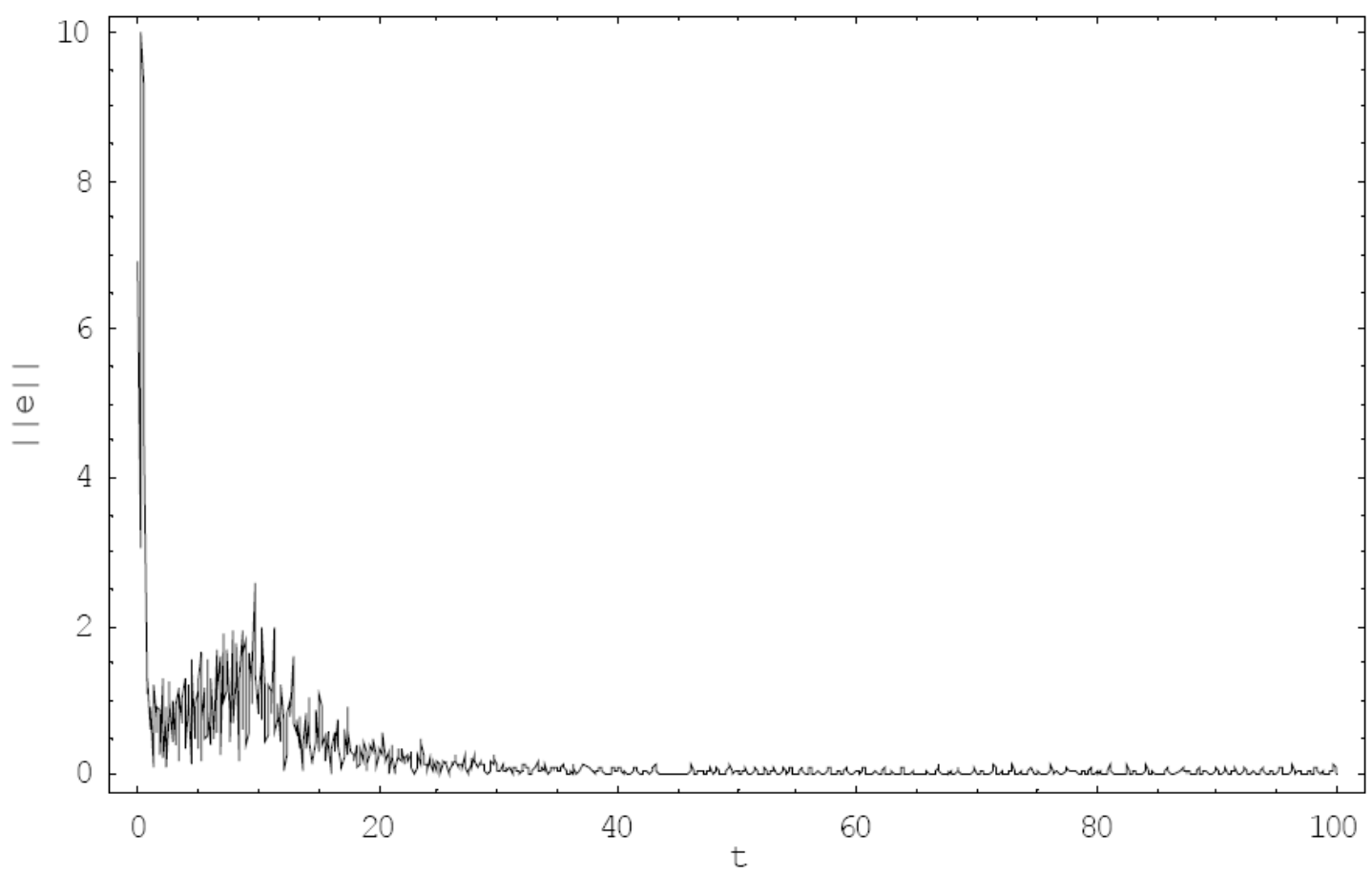

Fig.2 Achievement of synchronization with uniform ultimate bound under perturbation (12)

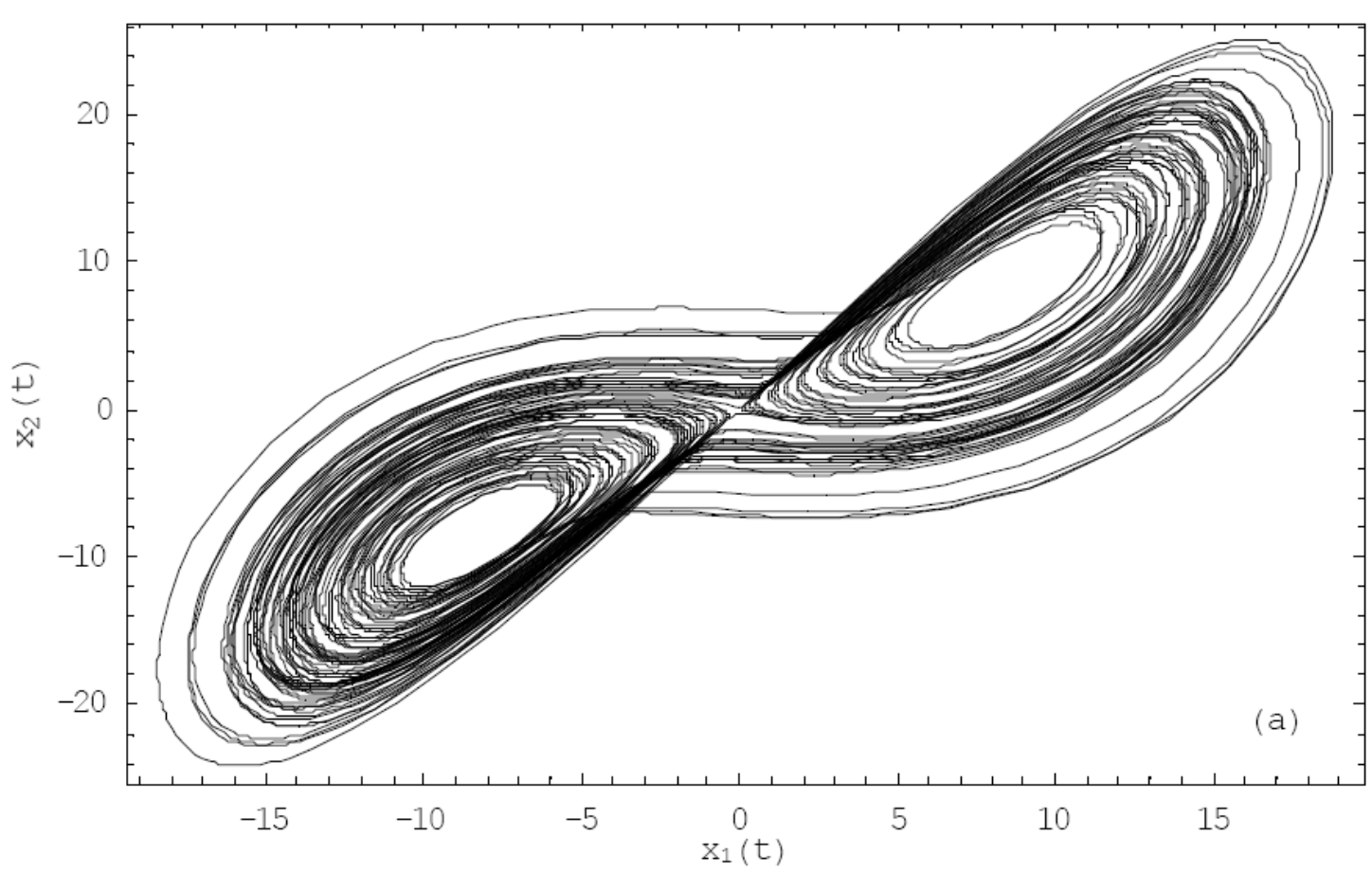

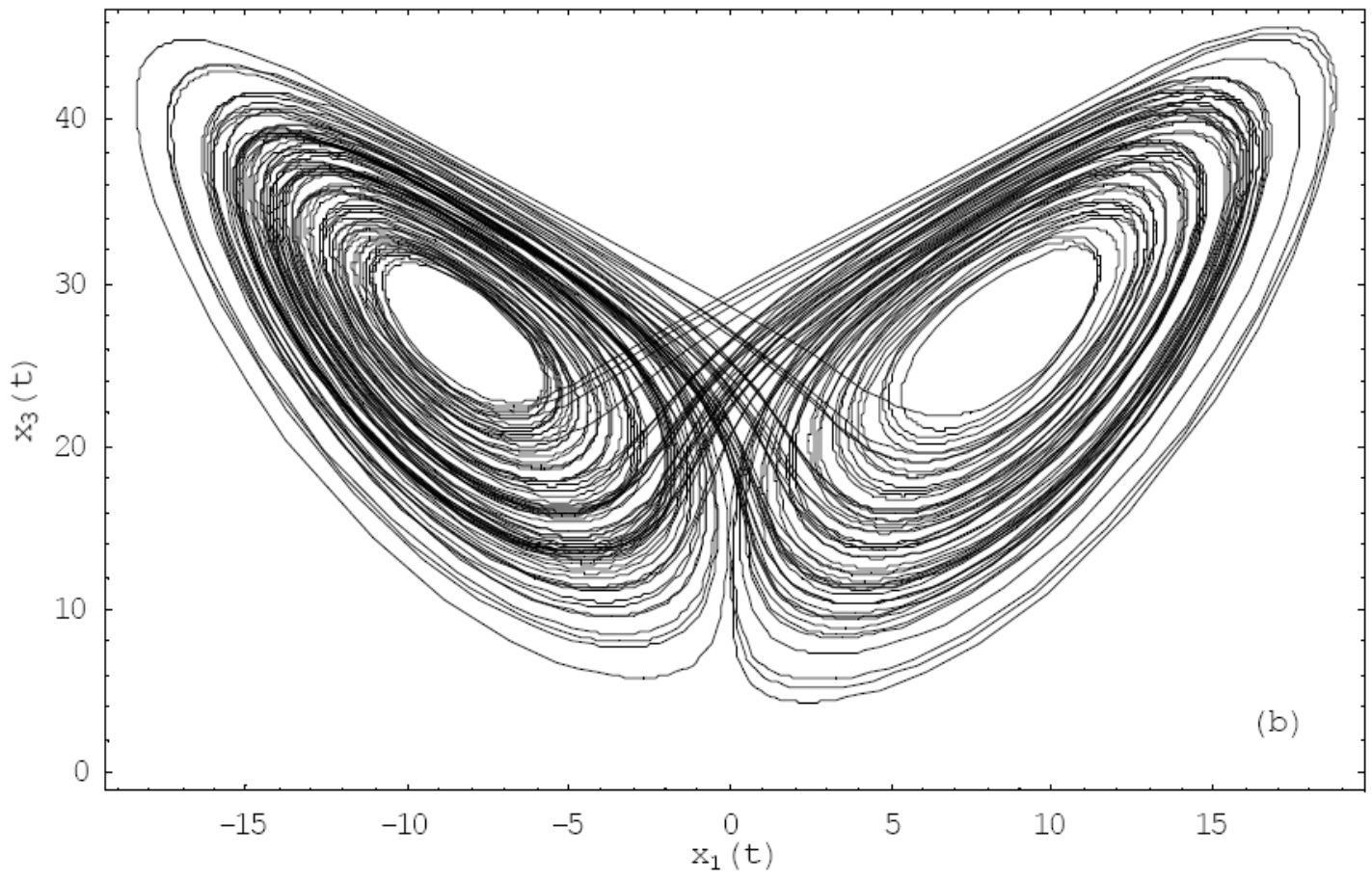

Fig.3 Attractor of the Lorenz system under perturbation (12): (a) $x_1 - x_2$ plane; (b) $x_1 - x_3$ plane

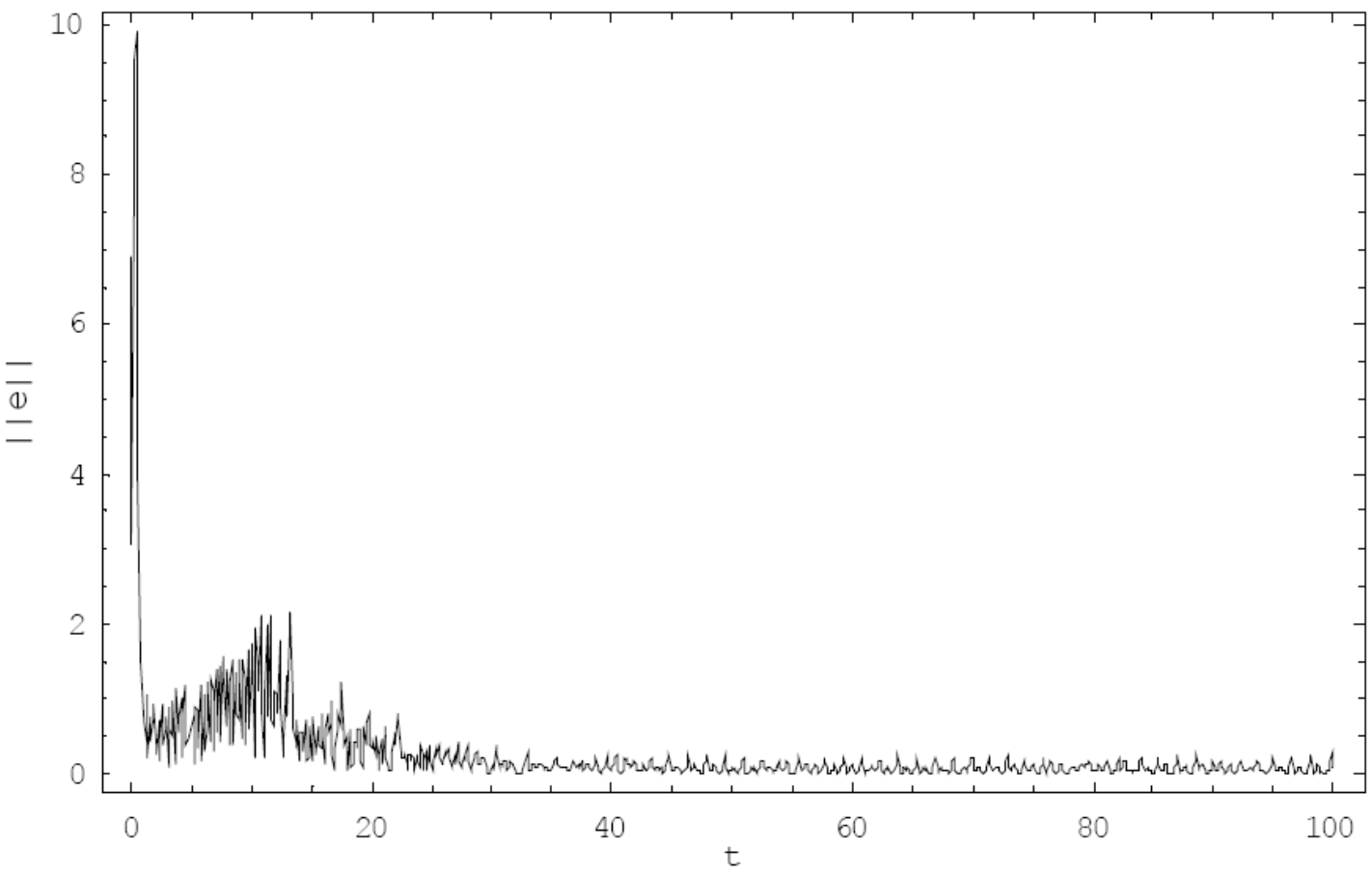

Fig.4 Synchronization error of drive-response systems (10) and (11) under perturbation (12) with strength $c = 10\%$

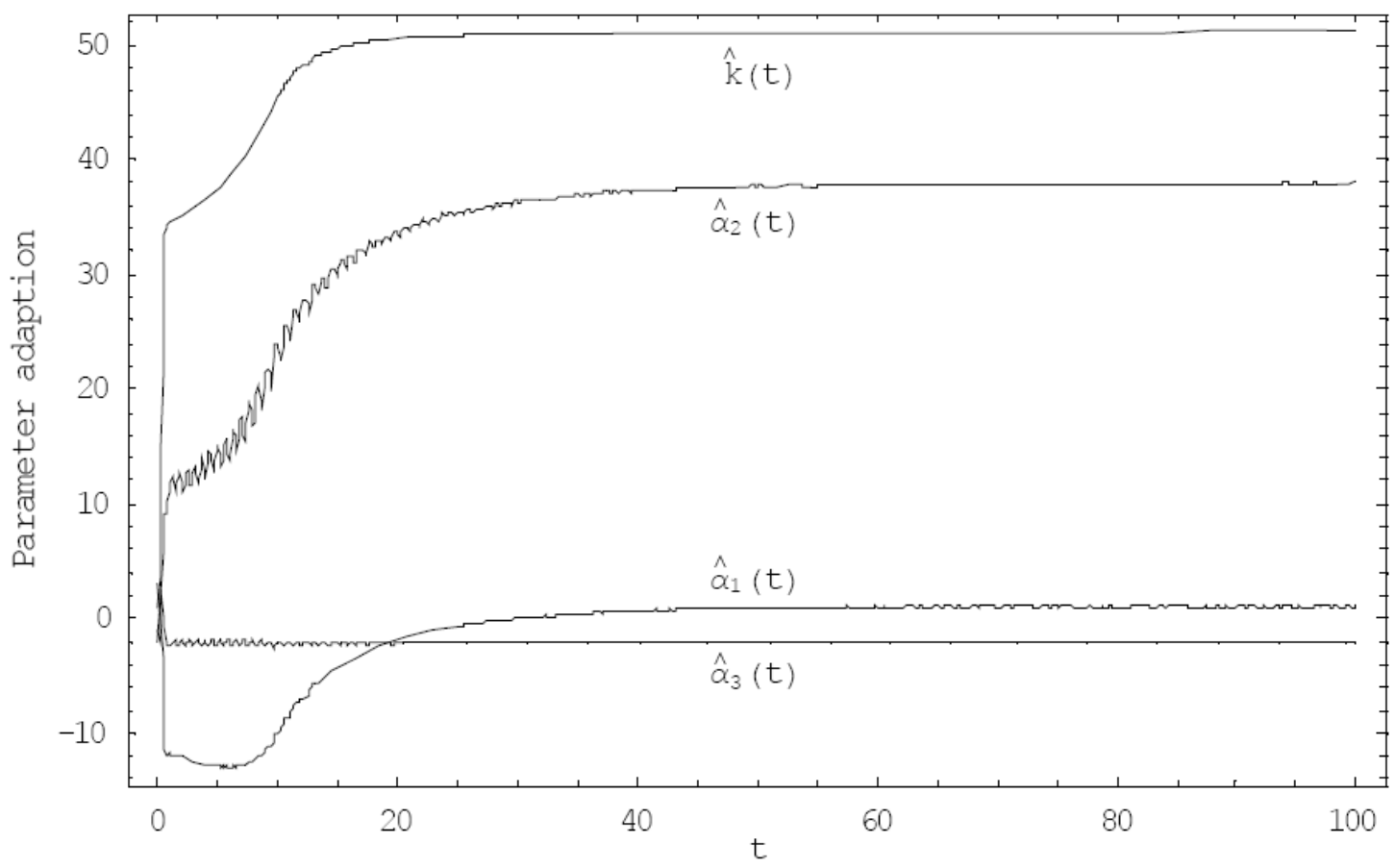

Fig.5 Adaptive processes of parameters $\hat{\alpha}_1$, $\hat{\alpha}_2$, $\hat{\alpha}_3$ and $\hat{k}$ under perturbation (12)

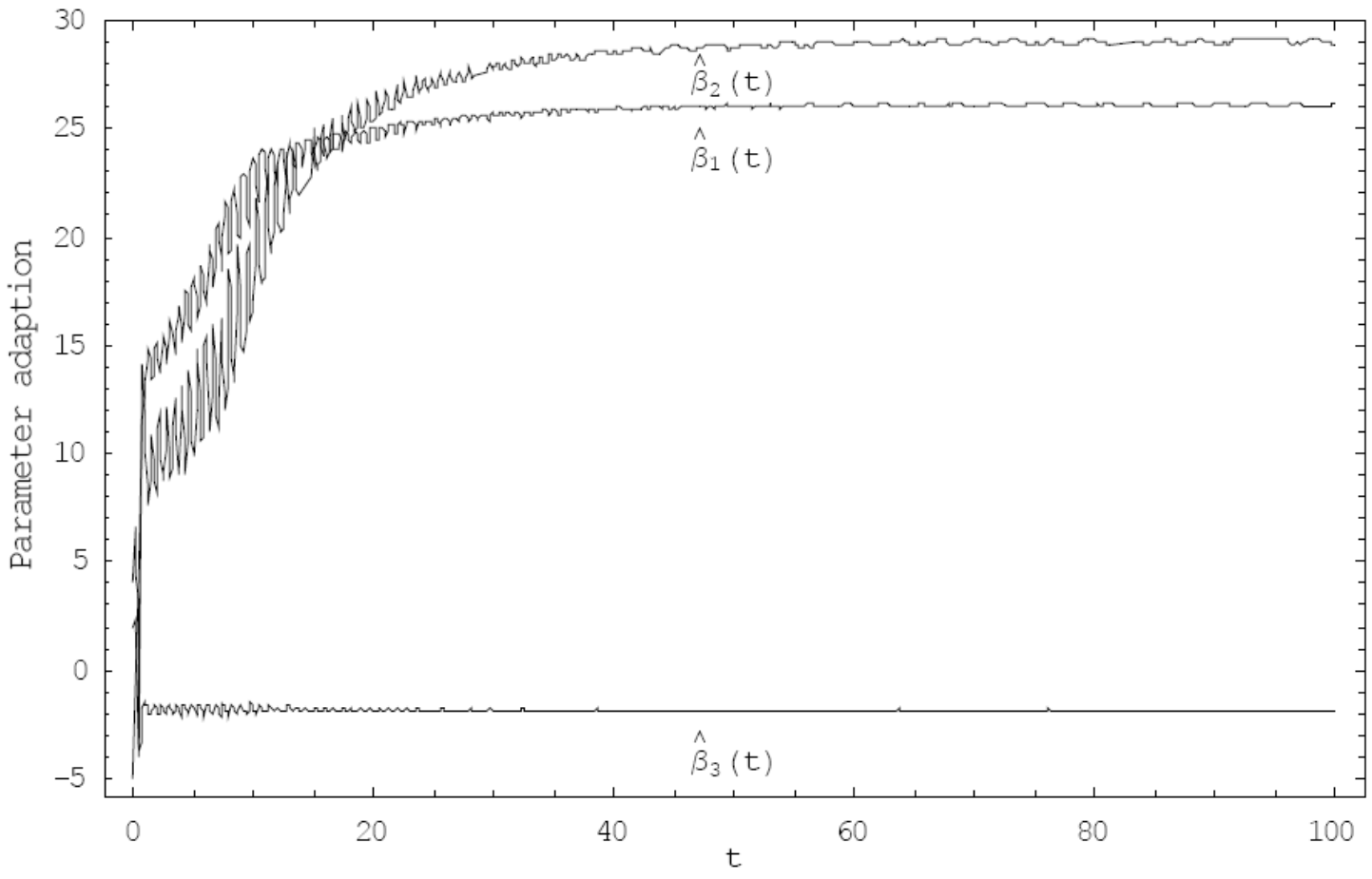

Fig.6 Adaptive processes of parameters $\hat{\beta}_1$, $\hat{\beta}_2$ and $\hat{\beta}_3$ under perturbation (12)

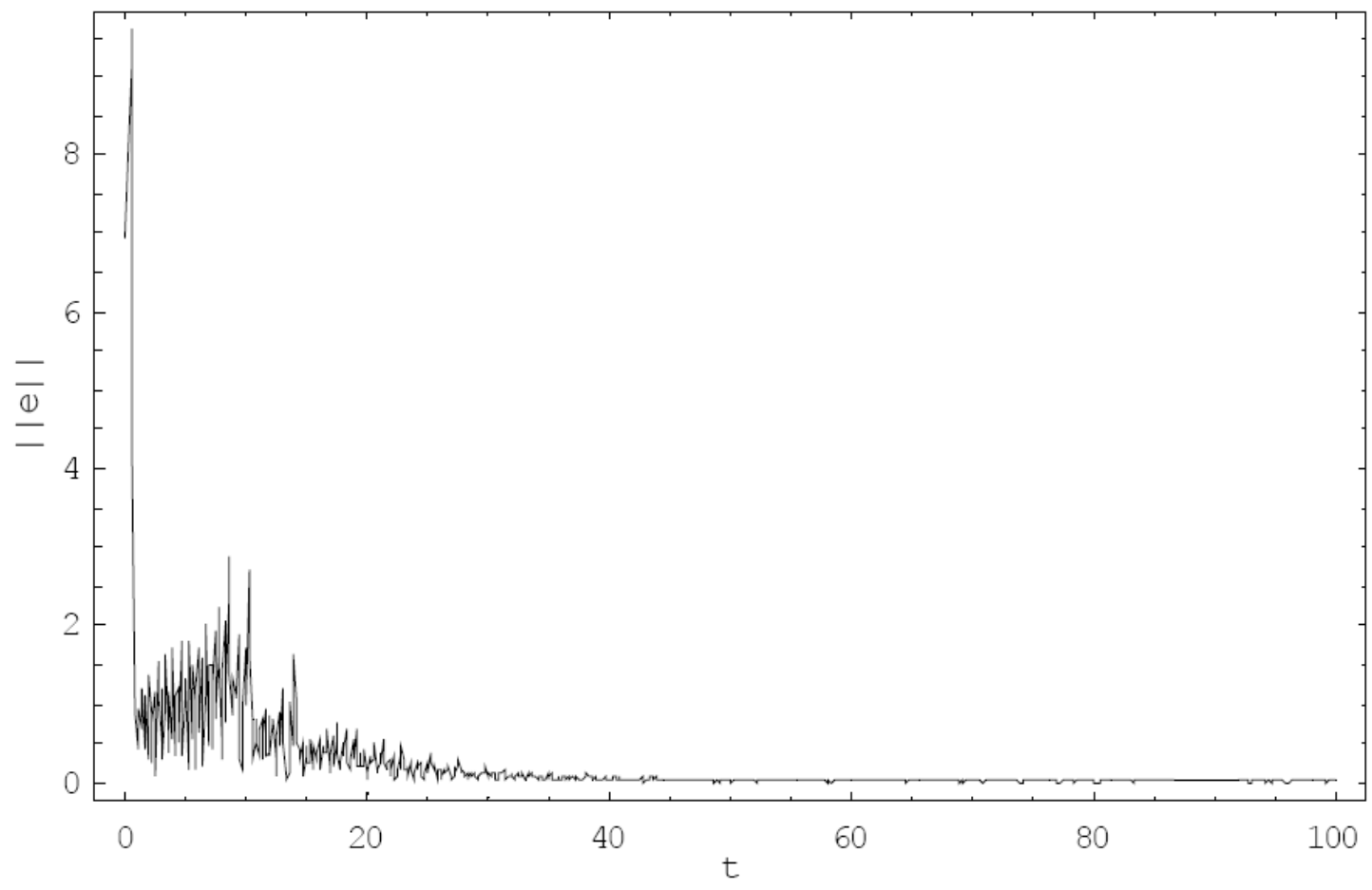

Fig.7 Achievement of synchronization with uniform ultimate bound under perturbation (13)

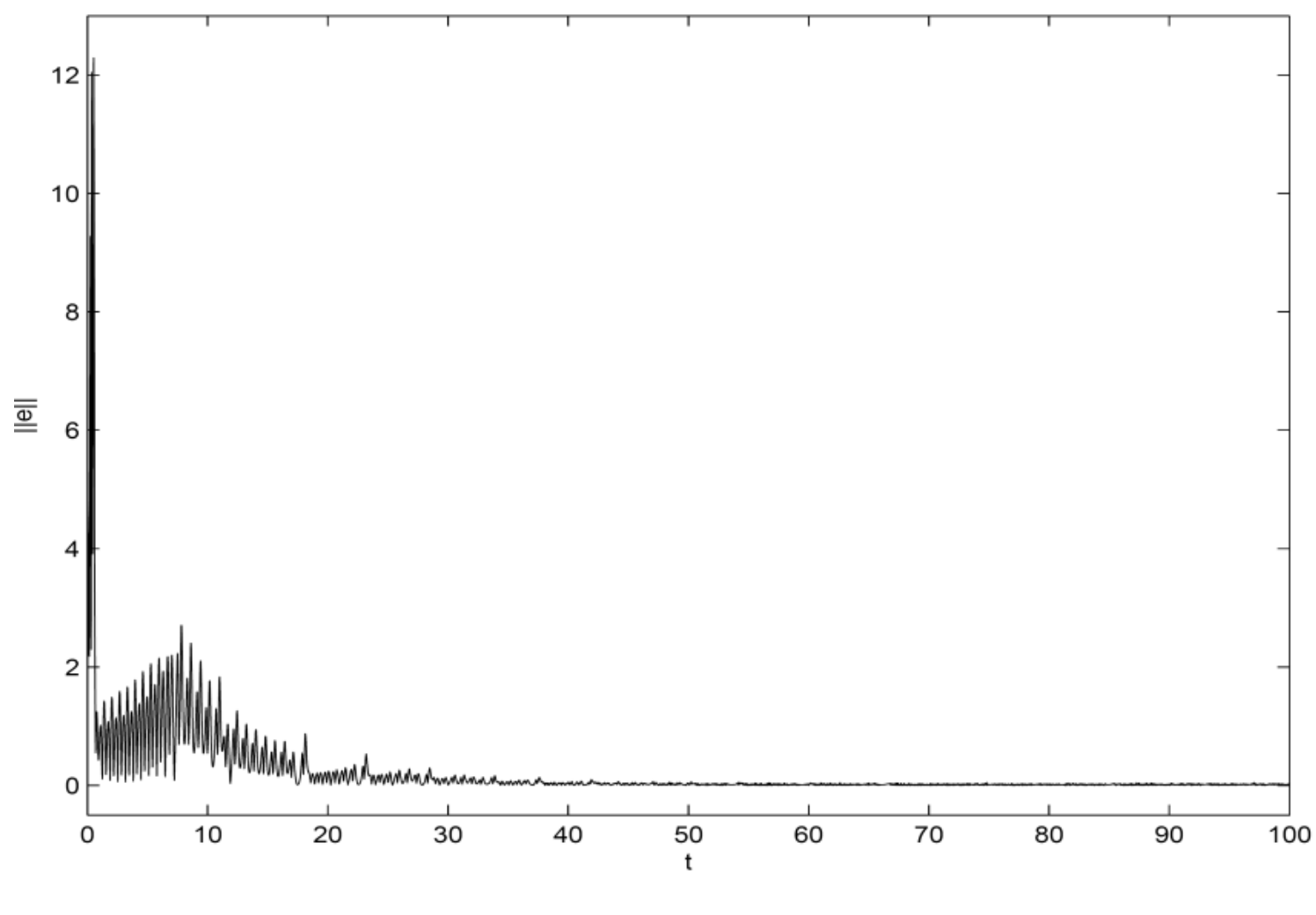

Fig.8 Achievement of synchronization with uniform ultimate bound under perturbation (14)